\documentclass[preprint,pre,floatfix,superscriptaddress]{revtex4}
\usepackage{graphicx}
\usepackage{epic,eepic}
\usepackage{graphics}
\usepackage{epsfig}
\usepackage{subfigure}

\newcommand{\be}{\begin{equation}}
\newcommand{\ee}{\end{equation}}
\newcommand{\bea}{\begin{eqnarray}}
\newcommand{\eea}{\end{eqnarray}}

\usepackage{amsmath}
\usepackage{amsfonts}
\usepackage{amssymb}
\usepackage{bm}
\usepackage{color}
\usepackage{graphicx}

\begin{document}
\title{Second order closure for stratified convection: bulk region
  and overshooting} 
\author{L. Biferale}\affiliation{Dept.
  Physics and INFN, Univ. of Tor Vergata Via della Ricerca
  Scientifica 1 00133 Rome Italy} 
\author{F. Mantovani}
\affiliation{Deutsches Elektronen-Synchrotron, Platanenallee 6, 15738
  Zeuthen, Germany} 
\author{M. Sbragaglia}\affiliation{Dept.
  Physics and INFN, Univ. of Tor Vergata Via della Ricerca
  Scientifica 1 00133 Rome Italy}
\author{A. Scagliarini}\affiliation{Dept.
  Physics and INFN, Univ. of Tor Vergata Via della Ricerca
  Scientifica 1 00133 Rome Italy}
\author{F. Toschi} \affiliation{Department of
  Physics and Department of Mathematics and Computer Science
  Eindhoven University of Technology 5600 MB
  Eindhoven The Netherlands\\
  and CNR-IAC Via dei Taurini 19 00185 Rome Italy}
\author{R. Tripiccione} \affiliation{Dipartimento di Fisica
  Universit\`a di Ferrara and INFN, via G. Saragat 1 44100
  Ferrara Italy}

\begin{abstract}
  The parameterization of small-scale turbulent fluctuations in convective
  systems and in the presence of strong stratification is a key issue for many
  applied problems in oceanography, atmospheric science and
  planetology. In the presence of stratification, one needs to cope
  with bulk   turbulent fluctuations and  with inversion
  regions, where temperature, density --or both-- develop highly
  nonlinear mean profiles due to the interactions between the
  turbulent boundary layer and the unmixed --stable-- flow above/below it. We
  present a second order closure able to cope {\it
    simultaneously} with both bulk and boundary layer regions, and we
  test it against high-resolution  
  state-of-the-art 2D  numerical simulations in a convective and
  stratified belt for values of the Rayleigh number, up to
  $Ra \sim 10^9$. Data are taken from a
  Rayleigh-Taylor system confined by the existence of an adiabatic
  gradient. 
\end{abstract}

\pacs{47.20.Ma,47.40.-x,42.68.Bz}

\maketitle
Realistic parameterization of convective regions in presence of strong
stratification is a problem of interest for the evolution of the
convective boundary layer in fields as different as atmospheric
science \cite{Siebesma07}, stellar convection
\cite{rt_atmosphere,Ludwig99}  and  oceanography 
\cite{Large94,Burchard01,Canuto,Wirth}.
The problem is intriguing also from a pure theoretical point of view: one
would like to predict the interplay between buoyancy and turbulence at the edge
between a convective region and a stable stratified volume above/below
it. Due to turbulent activity, intermittent puffs of
temperature, density and momentum tend to enter the stably stratified
region, locally producing an inversion in the energy balance:
kinetic energy is indeed lost, as potential energy is increased. 
Those turbulent puffs are the result of intense plumes
traveling across the   volume, often generated in the
bottom layer, producing a ``non-local'' interaction between edge
and bulk of the turbulent medium.
The problem, already recognized in the early 50's
\cite{Vitense53,spiegel3}, is still a subject of intense research
(see e.g. \cite{Burchard01}). 
Recent studies have shown that, in the absence of strong
stratification, mixing length theory based on Prandtl or Spiegel
closure works very well in situations as different as for the case of
Rayleigh-Taylor systems \cite{boffetta.prl.2010} or in fluid mixing by
gravity currents \cite{Ecke09}. 
\begin{figure}[!h]
\begin{center}
  \includegraphics[scale=0.2]{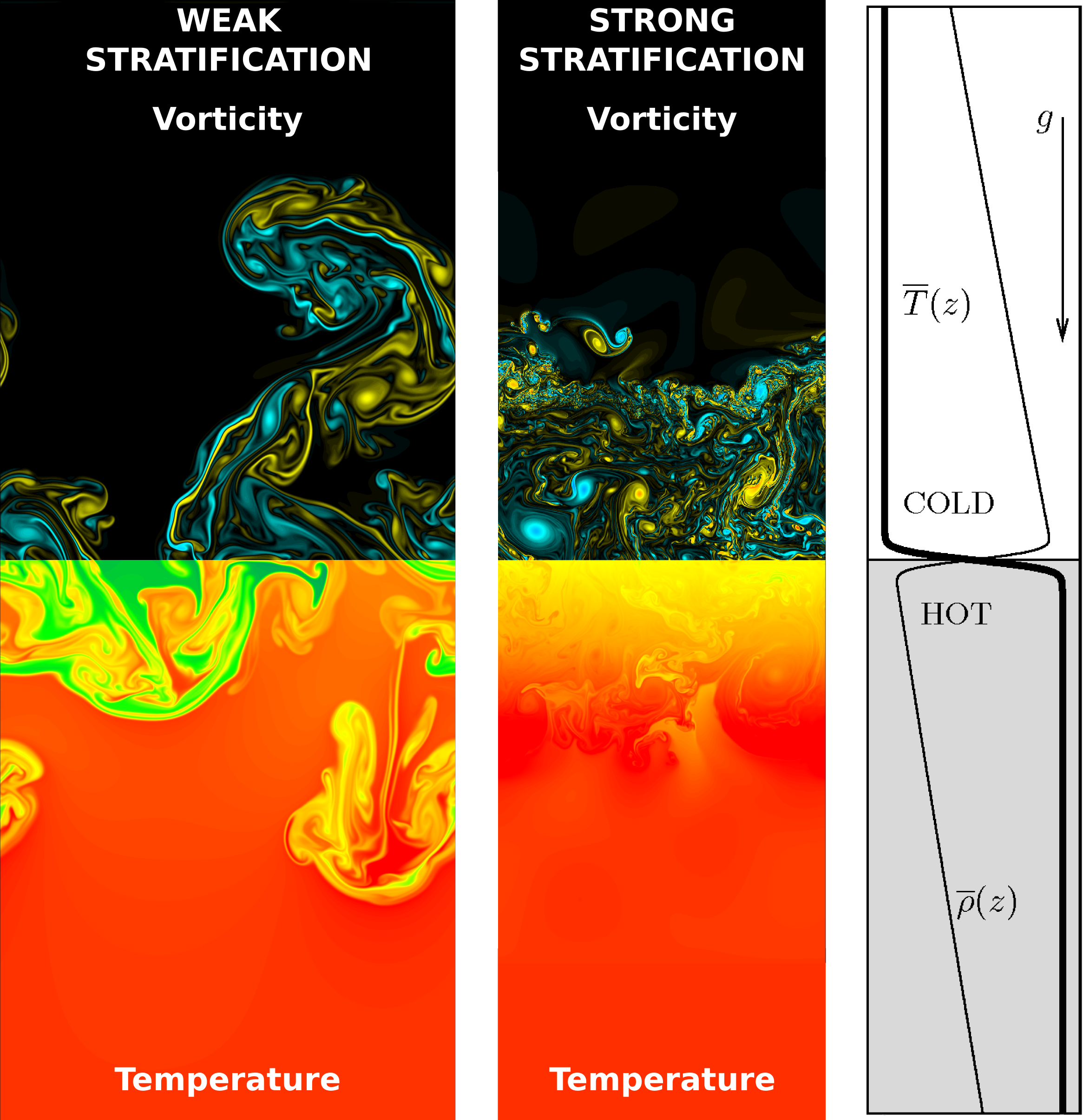}
  \caption{Right panel: initial vertical density and temperature profiles. Left and middle panels: Vorticity and temperature snapshots with weak
 and strong  stratification, respectively.  Notice the free rising  plumes 
 in the left panel typical of non-stratified RT
    system. In the middle panel stratification is
    stronger and turbulence is confined below and above the unmixed fluid at rest: at the boundary the temperature profile overshoots.}
\label{fig:v_e}
\end{center}
\end{figure}
Nevertheless, whenever strong stratification stops the evolution of
the mixing profile, an overshooting region with temperature inversion
develops and local closures fail.  The problem arises from non-local
effects caused by turbulent plumes or convective updraft.

\begin{figure}[h]
\begin{center}
  \advance\leftskip-0.15cm
  \includegraphics[scale=0.7]{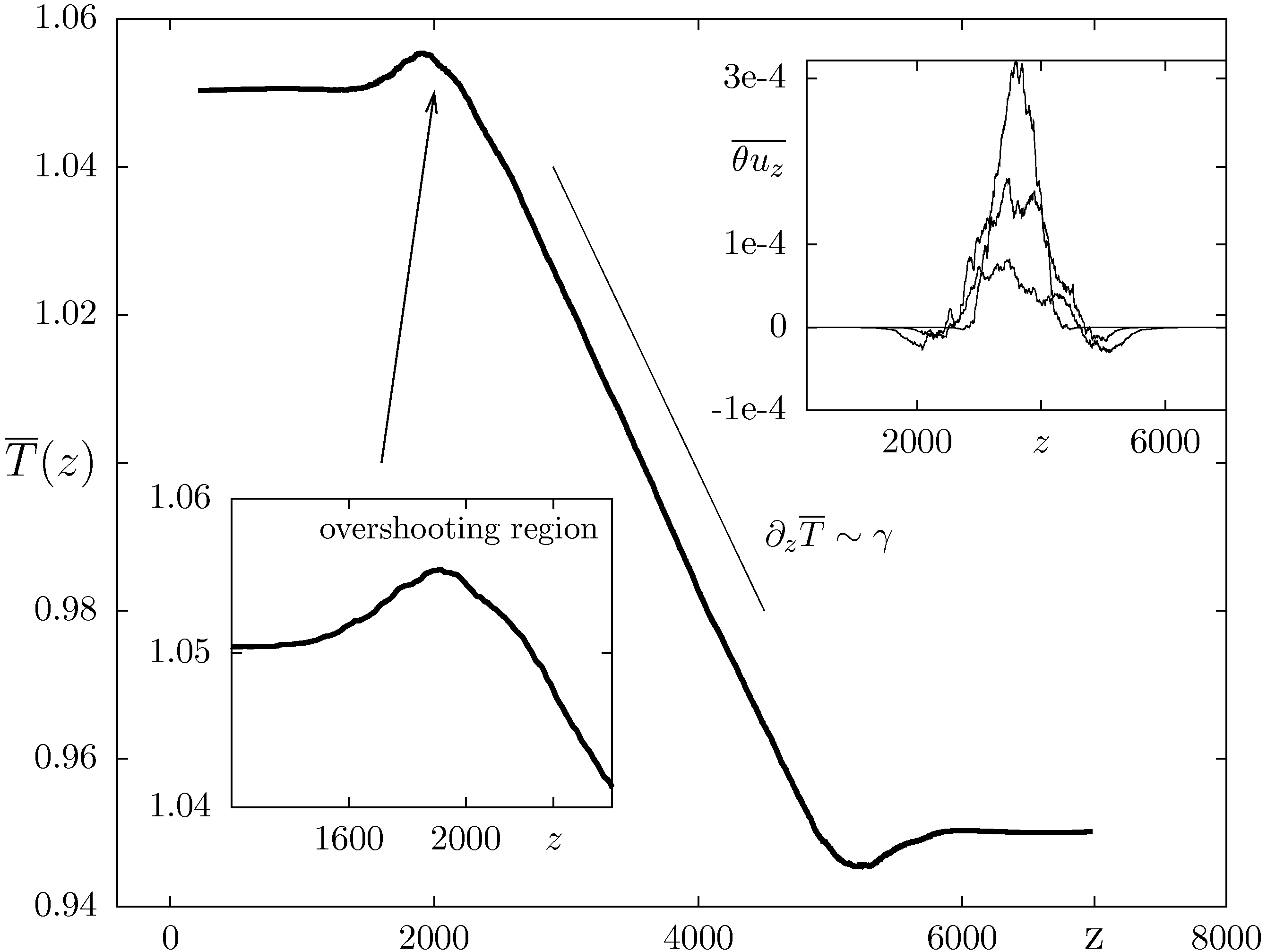}
  \advance\leftskip-0.55cm
  \caption{RUN (B). Mean temperature profile $\overline{T}(z)$ after the RT
    evolution has stopped at the adiabatic slope (solid line). 
Bottom left panel: zoom of the
    overshooting region, with temperature inversion. Top
    right inset: heat flux profile $\overline{u_z \theta}(z)$ at
    three times before, during and after stopping: two
    regions develp with negative heat flux in correspondence of the
    temperature overshooting.}
\label{fig:1}
\end{center}
\end{figure}

The situation can be visualized in the middle panel of
Fig. \ref{fig:v_e} where a highly turbulent configuration is confined
by two stably stratified media below and above it, at an effective
Rayleigh number, $Ra \sim 10^9$. The mean temperature profile is
linear in the bulk and it develops two --symmetric-- overshooting
regions at the edge between the turbulent boundary layer and the fluid
at rest with the heat flux inverting sign (see Fig. \ref{fig:1}).  The
configuration is obtained by evolving a Rayleigh-Taylor (RT) system in
2D with stratification
\cite{chertkov,celani1,pof1,detlef_arfm,rt.temp,cabot}.  The evolution
of the RT mixing layer, unbounded in the absence of stratification, is
stopped because of the adiabatic gradient, i.e.  when the mean profile
of the potential temperature becomes flat.  In this letter we propose
a closure for the evolution of the mixing layer, able to capture both
the initial transient (free of any stratification effect) and the late
slowing down and stopping due to stratification effects. The main idea
is to go beyond single point closure for the mean temperature
evolution,
keeping it{\it exact} and closing only for second order quantities:
the Reynolds stresses, the heat flux and  temperature fluctuations, see e.g.  \cite{Burchard01}.
We test our closure against state-of-the-art numerical simulations at
high resolution.  We choose to work in a 2D geometry to maximize the
capability to get quantitative measurements at high Reynolds and
Rayleigh numbers.  Choosing a RT system presents the extra complexity
of non-stationarity, allowing to probe also turbulent time scales.
Our numerical simulations are performed using a recently proposed
thermal Lattice Boltzmann Method \cite{JFM.nostro} able to reproduce
the Navier-Stokes equations for momentum, density and internal energy.
Validation of the method can be found in \cite{pof1}. Here we present
numerical simulations up to $ 4096 \times 10000$ grid points. Table I
provides details of our numerical experiment, obtained running on the
QPACE supercomputer \cite{qpace1}.
We limit our discussion here to the case of an ideal gas, with the
equation of state $P=\rho T$, and at small Mach number.  In this case,
the main effect of stratification is limited to the presence of an
adiabatic gradient affecting the evolution of temperature
\cite{spiegel3}. The equations are the following: (double indexes are
summed): \be
\begin{cases}
  \label{1.a}
  \partial_t u_i  + u_j \partial_j u_i = - \frac{\partial_{i} p}{\rho_m} -
 \delta_{i,z} g \frac{\theta}{T_m} + \nu
 \partial_{jj} u_i;\qquad i=x,z  \\
  \partial_t T + u_j \partial_j T -u_z \gamma = \kappa \partial_{jj} T ; \\
\end{cases}
\ee 
where $p$ is the deviation of pressure from the hydrostatic
profile, $p = P-P_H$ and $\partial_z P_H = -g \rho_H$, $T_m$ and
$\rho_m$ are the mean temperature and density in the system, $g$ is gravity and the
adiabatic gradient is given by its ideal gas expression $\gamma =
-g/c_p$ with $c_p$ the specific heat.  
\begin{table}
  \begin{center}
    \begin{tabular}{|c | c c c c c c c|}
      \hline   & $L_x$  & $L_z$   & $\gamma$ &   $Ra_{max}$ & $N_{conf}$ & $L_{\gamma}$  & $t_{RT}$\\
      \hline (A) & $4096$ & $10000$ & $-1\cdot 10^{-5}$    & $8\cdot 10^9$     & $18$      & $10000$ & $6.4 \cdot 10^4$ \\
      \hline (B) & $3072$ & $7200$ & $-4.2 \cdot 10^{-5}$   & $3 \cdot  10^9 $     & $11$   & $2408$ & $2.7 \cdot 10^4$ \\
      \hline
    \end{tabular}
    \caption{Run (A): weak stratification. Run (B) strong
      stratification.  Adiabatic gradient: $\gamma =-g/c_p$; $c_p=2$.
      Adiabatic length in grid units: $L_{\gamma} = \Delta T
      /|\gamma|$, $\Delta T = T_{down}-T_{up}$, $T_m =
      (T_{down}+T_{up})/2$ and $T_{up}=0.95$,
      $T_{down}=1.05$. Rayleigh number in presence of stratification
      is defined as \cite{spiegel3}: $ Ra(t) = g L(t)^4 (\Delta T/L(t)
      +\gamma)/(\nu\, \kappa)$). Maximal value is obtained when $L(t)
      = 3/4 L_{\gamma}$. Number of independent runs with random
      initial perturbation: $N_{conf}$. Atwood number = $At = \Delta
      T/(2\,T_m) = 0.05$. Characteristic time scale, $t_{RT} =
      \sqrt{L_x/(g At)}$.  }
\label{table:param} 
\end{center}
\end{table}
In this {\it Boussinesq}
approximation \cite{spiegel3} for stratified flows, momentum is forced only by
temperature fluctuations $ \theta = T -\bar T$ where we use the symbol
$\overline{(\cdot)}$ to indicate an average over the
horizontal statistically-homogeneous direction.  
\begin{figure}[h]
\begin{center}
  \advance\leftskip-0.15cm
  \includegraphics[scale=0.3]{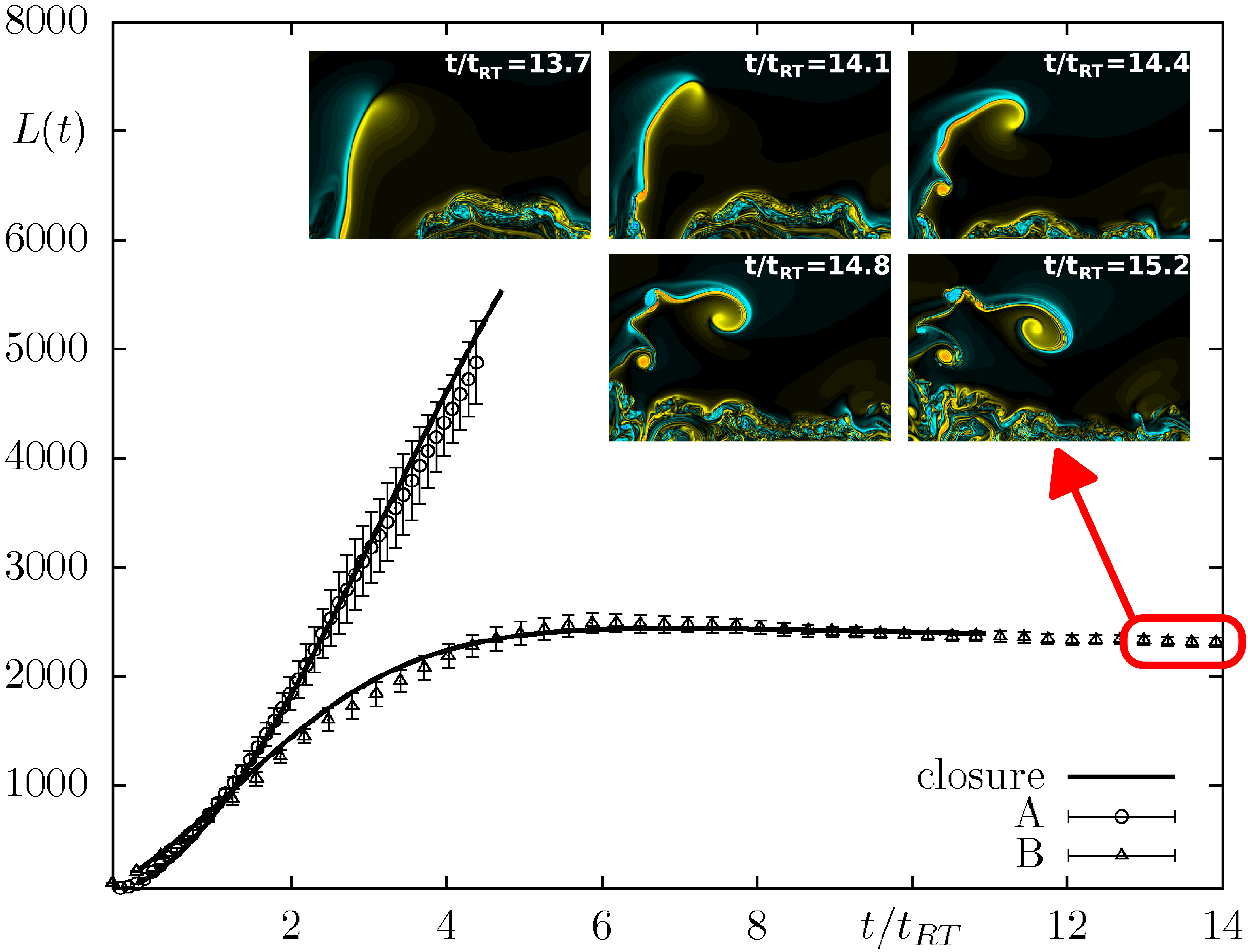}
  \advance\leftskip-0.55cm
  \caption{Evolution of the mixing layer extension $L(t)$ for both run
    (A) and (B).  Notice the stop by stratification effects for the
    latter.  Solid lines correspond to the closure prediction for the
    mixing length evolution obtained with the choice $[1.45,4.5]$ (run
    A) and $[0.3,6.5]$ (run B) for the parameters
    $[b_{\theta},b_{\theta u_z}]$ (see text). We also show 5
    consecutive snapshots of the overshooting region highlighting the
    formation of a turbulent plume trying to enter the stable region
    and rejected back by gravitational forces.}
  \label{fig:0}
\end{center}
\end{figure}
The initial configuration is given by two regions of cold (top) and
hot (bottom) fluids prepared in the two half volumes of our cell (see
right panel of Fig.\ref{fig:v_e}); turbulence is triggered by a small
perturbation of the interface between them \cite{rt.review} (see
\cite{boffi} for a recent high resolution
study of the classical RT system in 3D).\\
From (\ref{1.a}), one easily derives the equations for the mean
temperature profile, total kinetic energy, heat flux and temperature
fluctuations: \be
\begin{cases}
\label{eq:main}
  \partial_t \overline{T}+ \partial_z \overline{u_z \theta} =  \kappa \partial_{zz} \overline{T}\\
  \frac{1}{2} \partial_t \overline{\theta^2} + \frac{1}{2}\partial_z \overline{\theta^2 u_z}
  + \overline{u_z \theta} (\partial_z \bar T - \gamma) = \kappa
  \overline{\theta \partial_{jj} \theta} \\
\frac{1}{2} \partial_t \overline{u^2} +\partial_z [\overline{u^2 u_z} +\overline{u_z p}]= -g \overline{\theta u_z} -\epsilon_\nu \\
\partial_t \overline{\theta u_z} +\partial_z [\overline{\theta u_z^2} +\overline{\theta p}]= -g \overline{\theta^2} +\beta(z)\overline{u_z^2} -\epsilon_{\theta,u_z} 
\end{cases}
\ee where $u^2=u_x^2+u_z^2$, $\beta(z) = (\partial_z \overline{T}
-\gamma)$, and the dissipative terms are: $\epsilon_{\theta,u_z} =(\nu
+ \kappa) \overline{\partial_i \theta \partial_i u_z}$,
$\epsilon_{\theta}= \kappa \overline{\partial_i \theta \partial_i
  \theta}$, $\epsilon_{\nu} = \nu\overline{\partial_i u_j \partial_i
  u_j}$. The above set of equations is exact, except for boundary
dissipative contributions as for example, $ \kappa \partial_z
\overline{\theta \partial_z \theta}$ which are irrelevant when
$\kappa,\nu \rightarrow 0$ in absence of walls.  From the second of
(\ref{eq:main}) it is evident that temperature fluctuations are not
forced anymore as soon as the mixing region develops a mean
temperature profile of the order of the adiabatic gradient: \be
\partial_z \overline{T} \sim \gamma.  \ee As a consequence, from that
time on, turbulence will decline.  Given $\gamma$, we can identify the
adiabatic length, $L_{\gamma} = \Delta T/|\gamma|$ which corresponds
to the prediction for the largest possible extension of the mixing
layer during the RT evolution.  In Fig. \ref{fig:0} we show the
evolution of the mixing layer extension for two cases with weak (RUN
A) and strong (RUN B) adiabatic gradient. Clearly, the extension of
the mixing region stops to grow when $ L(t) \sim L_{\gamma}$. In this
paper we measure the mixing layer width $L(t)$ as in \cite{cabot}: \be
\label{eq:lcook}
L(t) = 2 \int dz\, \Theta \left[
  \frac{\overline{T}(z,t)-T_{up}}{T_{down}-T_{up}} \right], 
\ee 
with $\Theta[\xi] = 2 \xi;\;0 \le \xi \le 1/2$ and $ \Theta[\xi] = 2\,(1-\xi);\; 1/2 \le \xi \le 1 $.  
The overshoot developing at the edge between turbulent and unmixed
fluid is visible in Fig. \ref{fig:1}, where we show both the nonlinear
temperature inversion (inset bottom left) and the corresponding
inversion in the heat flux (inset top right). This overshooting region
is clearly a problem for any attempt to close the mean profile
evolution with any sort of {\it local} eddy diffusivity:
\begin{figure}
  \begin{center}
    \advance\leftskip-0.15cm
    \includegraphics[scale=0.7]{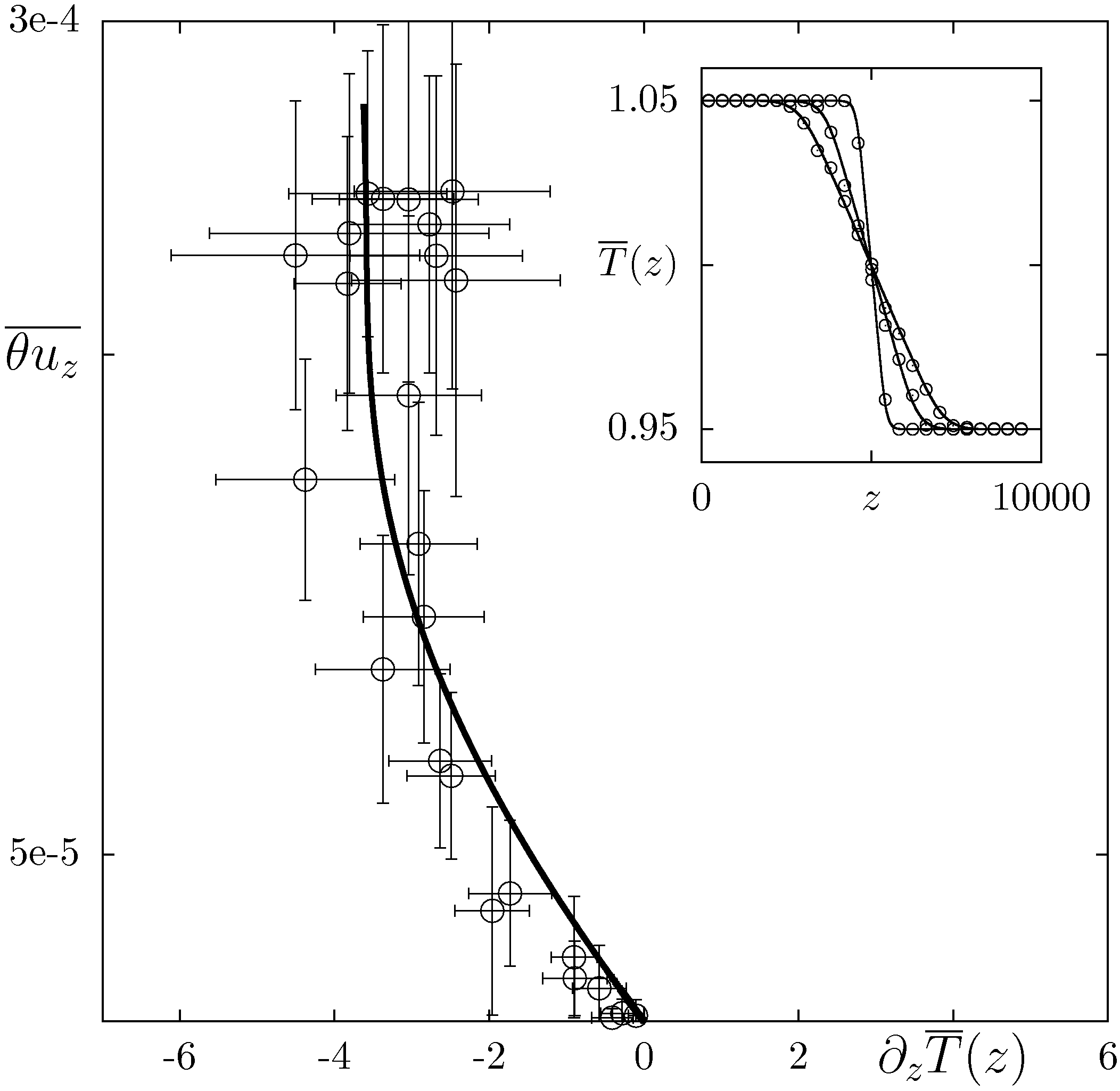}
    \advance\leftskip-0.55cm
    \caption{Check of the closure for the case with weak
      stratification (run A).  Circles: numerical data, solid lines:
      clousre.  We plot the local heat flux vs the local temperature
      mean gradient, $ \overline{\theta u_z} $ vs $(\partial_z
      \overline{T})\times 10^{5}$, i.e. the effective diffusivity
      $K(z,t)$ ().  Error bars are evaluated averaging over the
      configurations specified in table (1).  Result from the closure
      is given by the solid line with $[b_{\theta},b_{\theta
        u_z}]=[1.45,4.5]$ and $b_{u_z}=0.03$.  Inset: superposition of
      temperature profiles at three different times,
      $t/t_{RT}=2,4,5.5$ with the corresponding curves from the
      closure.}
    \label{fig:4}
  \end{center}
\end{figure}
$$
\overline{u_z \theta } = - K(z,t) \partial_z \overline{T}.
$$
Different models have been proposed in the literature for $K(z,t)$,
going from simple homogeneous closure ($K(z,t) \propto L(t) |\dot L(t)|$) to
Prandtl-like mixing-length closure \cite{boffetta.prl.2010,Ecke09}
($K(z,t) \propto L(t)^{5/2} \partial_z \overline{T}$) or following
Spiegel's nonlinear approach \cite{spiegel3} ($K(z,t) \propto L(t)^2
|\partial_z \bar T|^{1/2}$).  All these attempts work well in the absence of
overshooting and all of them suffer whenever an inversion in
temperature and heat flux is observed (as in Fig. \ref{fig:1}), 
implying a negative effective eddy diffusivity. 
In order to overcome this
difficulty, we keep exact the equations for the mean profile and
close only the fluctuations at the second order moments in (\ref{eq:main}). 
\begin{figure}
  \begin{center}
    \advance\leftskip-0.15cm
    \includegraphics[scale=0.7]{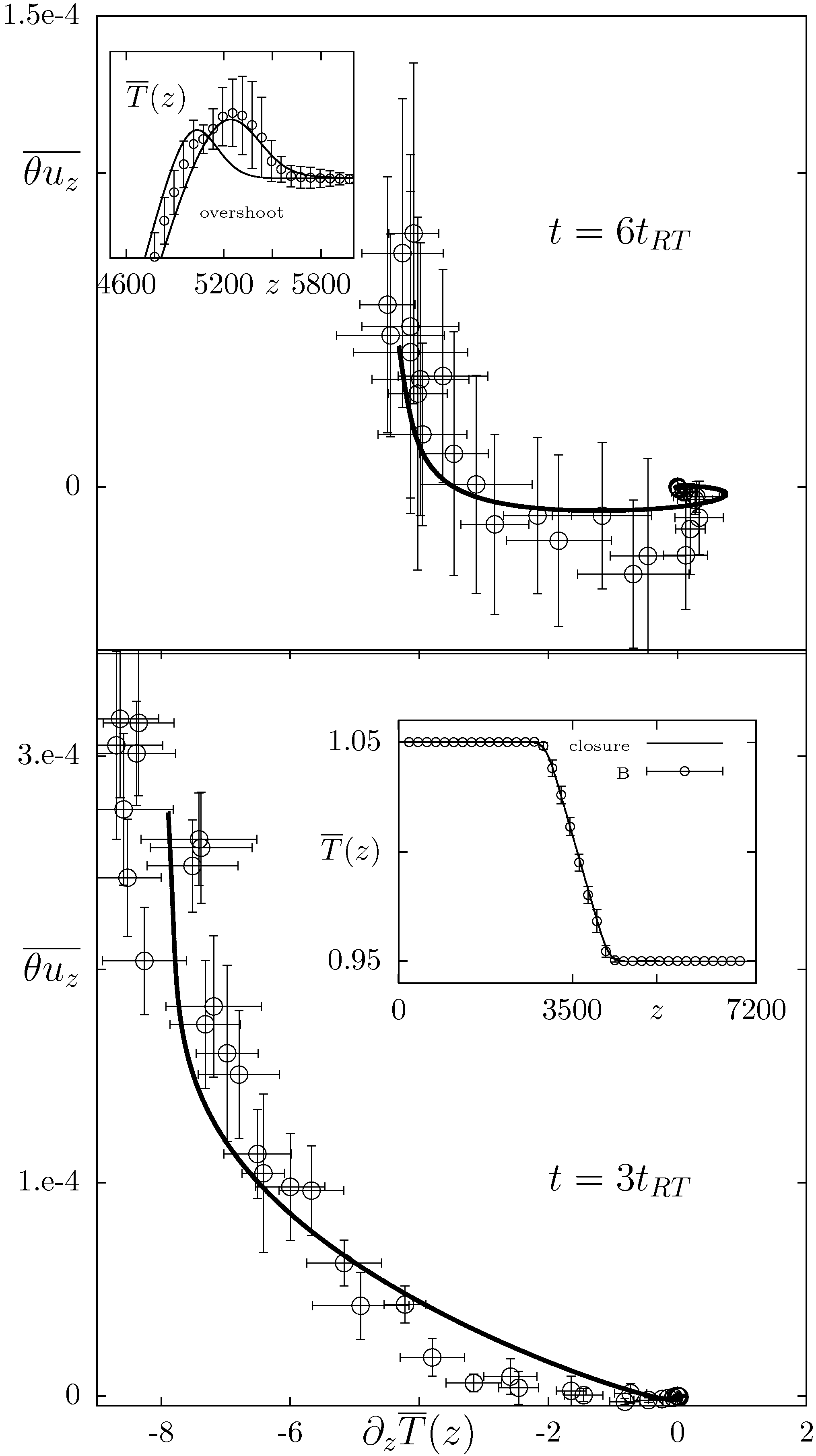}
    \advance\leftskip-0.55cm
    \caption{Check of the closure for the case with
      strong stratification (run B). Circles: numerical data,solid line: closure.We plot the local heat flux vs
      the local temperature mean gradient, $ \overline{\theta u_z}$ vs
      $(\partial_z \overline{T})\times 10^5$, i.e. the effective diffusivity
      $K(z,t)$ for two different times, $t= 3\, t_{RT}$ (bottom  panel);
      $t=6\, t_{RT}$ (top panel).
      Closure predictions ($[b_{\theta},b_{\theta u_z}] = [0.35;6.5]$)
      are given by the solid lines. 
Inset bottom panel: matching between the temperature profile and the closure. Inset  top panel: overshooting
      region around the top boundary
      layer. The two lines correspond to the closure
      using two different choices,
      $[b_{\theta},b_{\theta u_z}] = [0.25;4.5];[0.35,6.5]$.}
\label{fig:3}
\end{center}
\end{figure}
Considering  $u_z \sim u$,  we are left with six unknown to be 
defined: the three dissipative contributions on
the rhs, and the three non-linear third order fluxes on the lhs. We
close them adopting the simplest dimensionally-consistent {\it local}
closure, for both  fluxes and dissipative terms:
\be
\begin{cases}
\label{eq:eq1}
\overline{\theta^2 u_z}= a_{\theta} L |\dot L| \partial_{zz} \overline{\theta^2}; \qquad  \qquad \qquad  \epsilon_{\theta} = b_{\theta} { \overline{\theta^2}}/{\tau(z,t)} \\
\overline{(u^2+p)u_z} =a_{u_z}  L |\dot L| \partial_{zz} \overline{u_z^2}; \qquad \epsilon_\nu =b_{u_z} {\overline{u_z^2}}/{\tau(z,t)}\\
\overline{\theta (u_z^2 + p)}=a_{\theta u_z} L |\dot L| \partial_{zz}
\overline{\theta u_z} ;\qquad \epsilon_{\theta,u_z} = b_{\theta u_z} {
  \overline{\theta u_z}}/{\tau(z,t)} \nonumber \\
\end{cases}
\ee where the typical time defining the dissipation rates is fixed by
$\tau(z,t) = \sqrt{ \overline{u_z^2}}/L(t)$.  Some comments are in
order. First, we notice that the closure is now {\it local} but on the
second order moments, i.e. {\it non-local} for the evolution of mean
profiles.  Furthermore, out of the 6 free parameters, 4 can be handled
quite robustly, all the three coefficients in front of the closure for
third order quantities are set ${\cal O}(1)$ using a first order guess
from the numerics, $a_{\theta}=0.2; a_{u_z} = 0.3; a_{\theta u_z} =
0.8$. Moreover, the free parameter defining the intensity of the
kinetic dissipation, $\epsilon_\nu$, is irrelevant in 2D (absence of
direct energy cascade). The only two {\it delicate} free parameters
are those defining the intensities of temperature and heat-flux
dissipative terms, $[b_{\theta}, b_{\theta u_z}]$. In order to get a
good agreement with the numerics we need some fine tuning for them. It
is important to notice that both dissipative terms will develop a
non-trivial vertical profile, i.e. they are not given by a simple bulk
homogeneous parameterization.

In Fig. \ref{fig:0} we show that the
closure is able to reproduce quantitatively the evolution of $L(t)$ for both
cases of weakly stratified turbulence (A) and strongly stratified case (B). 
In Figs. (\ref{fig:4}-\ref{fig:3}) we show the capability of the model to
reproduce the heat flux {\it vs.} temperature profile behavior, providing a sort
of {\it aposteriori} effective eddy diffusivity. As one can see, the closure is
able to capture both the weak stratification case (Fig. \ref{fig:4}) and the
strong stratification case (Fig. \ref{fig:3}).  For the weak stratification case
the {\it aposteriori} eddy diffusivity is very close to the one predicted by using
the Prandtl mixing length theory as proposed in \cite{boffetta.prl.2010} or to
the one proposed by Spiegel's \cite{spiegel3} (not shown). For
the strong stratification case, our model is able to capture also the long time behavior, even after the
evolution has come to a halt due to the adiabatic gradient, where the
heat flux has completely inverted sign, see top panel of  Fig. \ref{fig:3}. 
In the inset of the same figure we also show the capability of the closure to
reproduce the overshooting profile. As one can see the agreement is very good and the results are
not very sensitive to the choice of the free parameters.\\
We also remark that from eqs. (\ref{eq:main}) one may derive a
dimensional closure neglecting all terms except the one at large
scales, and assuming that temperature fluctuations are dominated by
the global temperature jump: $\overline{\theta^2} \sim
\Delta T $. From the third equation we  derive:  $\overline{\theta u_z}
\sim L \overline{u_z^2}^{1/2} \partial_z \bar T $ and from the second:
 $\overline{u_z^2} \sim g L^2 \partial_z \bar T $ that --combined together--
leads to Spiegel's closure: $\overline{\theta u_z} \sim L^2 (\partial_z \bar
T)^{1/2} \partial_z \bar T $.

In conclusions, we have performed state-of-the-art 2D numerical
simulations using a novel LBM for turbulent convection driven by a
Rayleigh-Taylor instability in weakly and strongly stratified
atmospheres. For the strongly stratified case, we are able to resolve
the overshooting region with up to 800 grid points, something
impossible to achieve with 3D direct numerical simulations because of
lack of computing power.  We have presented a second-order closure to
describe the evolution of mean fields able to capture both bulk
properties and the overshooting observed at the boundary between the
stable and unstable regions. The closure is {\it local} in terms of
fluctuations while keeping the evolution of the mean temperature
profile exact.  In order to apply the same closure to 3D cases one
needs to take into account some possible non-trivial effects induced
by the kinetic energy dissipation modeling, due to the presence of a
forward energy cascade. As a result, also the parameter $b_{u_z}$ will
become a relevant input in the model.
 
We acknowledge useful discussions with G. Boffetta and A. Wirth. We
acknowledge access to QPACE and eQPACE during the bring-up phase of
these systems. Parts of the simulations
were also performed on CASPUR under HPC Grant 2009 and 2010. \\

\end{document}